\documentclass[a4paper,11pt]{article}
\pdfoutput=1 

\usepackage{jcappub} 

\usepackage[T1]{fontenc} 

\title{Vacuum non-linear electrodynamic polarization effects in hard emission of pulsars and magnetars}

\author[a]{V.I. Denisov,}
\author[a,1]{V.A.Sokolov,\note{Corresponding author.}}
\author[a]{S.I.Svertilov}

\affiliation[a]{Physics Department, Moscow State University, Moscow, Russia}

\emailAdd{vid.msu@yandex.ru}
\emailAdd{sokolov.sev@inbox.ru}
\emailAdd{sis@coronas.ru}

\abstract{The non-linear electrodynamics influence of pulsar magnetic field on the electromagnetic pulse polarization
is discussed from the point of observation interpretation. The calculations of pulsar magnetic field impact
 on electromagnetic pulse polarization are made in such a way to make
it easier to interpret these effects in space
  experiments. The law of hard emission pulse propagation in the pulsar
 magnetic field according to the vacuum
  non-linear electrodynamics is obtained. It was shown, that due to the
 birefringence in the vacuum the front
  part of any hard emission pulse coming from a pulsar should be
 linearly polarized and the rest of pulse can have
   arbitrary polarization. Observational possibilities of vacuum
 birefringence are discussed.  The estimations of
   detector parameters such as effective area, exposure time and
 necessity of polarization measurements with high
   accuracy are given. The combination of large area and extremely
 long exposure time gives the good opportunity
   to search the fine polarization effects like vacuum non-linear
 electrodynamics birefringence.}


\begin{document}
\maketitle
\flushbottom

\section{Introduction}
\label{sec:intro}

 Although about a few hundreds of pulsars are discovered in
hard X-rays and gamma-rays and their temporal and
spectral parameters were studied with good accuracy, there are no
sufficiently completed data about
polarization properties of these astrophysical objects. A very limited
number of space experiments, which were
able to study polarization in hard emission, were realized at all.
Mainly, there were observations of solar
flares, such as polarization measurements at Soviet and Russian
satellites InterCosmos 
\cite{c1} 
and
Coronas-F 
\cite{c2} 
and RHESSI space observatory,
which  give the principle possibility of
polarization detection of solar flare gamma-rays and cosmic
gamma-ray bursts (GRB) \cite{c3} 

Polarimetry is a powerful diagnostic of specific phenomena at work in
cosmic sources in the radio-wave and optical
energy bands, but very few results are available at high photon
energies: the only significant observation in the
X-gamma energy range, to date, is the measurement of a linear
polarization fraction of $P = 19 + 1\%$ of  the 2.6 keV
emission of the Crab nebula by a Bragg polarimeter on board OSO-8
\cite{c4}.
At higher energies,
hard-X-ray and soft-gamma-ray telescopes that have flown to space
in the past (for ex. COMPTEL CGRO, 
\cite{c5}) 
were not optimized for polarimetry, and their sensitivity to
polarization was poor. Presently active missions
INTEGRAL IBIS 
\cite[]{c6,c7,c8}, 
and SPI 
\cite[]{c9,c10}, 
have provided some improvement, with, in particular, mildly
significant measurements of $P = 28 + 6\%$ (from 130
to 440 keV), 
\cite{c11} 
and $P = 47^{+19}_{-13}  \%$ (from 200 to 800 keV), 
\cite{c12} 
for the Crab Nebula.

 A number of Compton polarimeter/telescope projects have been
developed, some of which also propose to record
 photon conversions to $e^+ e^-$  pairs. A variety of technologies
have been considered, such as scintillator arrays
 POGO 
 \cite{c13}, 
GRAPE \cite{c14}, 
POLAR \cite{c15}, 
 Si or Ge microstrip detectors  MEGA 
\cite{c16}, 
ASTROGAM  \cite{b2} 
 or combinations
of those (Si + LaBr$_3$ for GRIPS 
\cite{c17}, 
Si + CsI(Tl)
for TIGRE 
\cite{c18}, semiconductor pixel detectors CIPHER 
\cite{c19}, 
liquid xenon LXeGRIT 
\cite{c20} 
and gas mixture CF4 at 3 atm
\cite{c21}) 
time
projection chambers (TPC).

It is supposed that hard emission of astrophysical sources
contained strongly
magnetized  neutron stars should be more-less polarized.
It could be caused by geometry of emitting areas,
i.e. accretion disk (in the case of neutron star in double system)
or magnetic field configuration as well as
by highly anisotropy of primary electron beams. The other reason is
connected with influence on X and gamma
ray propagation of specific physical conditions in neutron star
 magnetosphere.

However, there are the other kind of physical processes, which can lead to polarization effects. It means the
influence of very strong magnetic fields on the X and gamma ray polarization. The critical value is so called
Schwinger field $B_q=m^2c^3/(e\hbar) =4.41\cdot 10^{13}$ G, above which the non-linear electrodynamic effects in
vacuum become significant. It is impossible to realize such fields in ground experiments, so the
astrophysical observations are the only way to test non-linear electrodynamic models. The strong magnetic
field is not so rare in the Universe.

Some gamma-ray pulsars are characterized by magnetic fields near $B_q$ value, but for so called magnetars it
could be even higher and reach about $10^{15}$ G. In such objects in the vicinity of strongly magnetized
neutron star there are favorable conditions for non-linear electrodynamic effects, in particular, vacuum
birefringence \cite[]{c22,c36,c37}. This effect may have different manifestations. One is the
dramatically increases the linear
polarization of the thermal radiation of the isolated neutron stars \cite{b1}, from a level of
a few per cent up to even 100 per cent, depending on the viewing geometry and the surface emission mechanism.
Currently, this vacuum non-linear electrodynamics prediction can be tested by measurement of a polarization
degree
 of isolated neutron star thermal emission.

First observations \cite{c52} of optical polarization from a thermally emitting isolated neutron star RX
J1856.5-3754 shown that a linear polarization degree of this star thermal emission is 16.43+5.26 per cent.

It is arguing in \cite{c52} that, independently on how thermal photons are produced, such a high value of
linear polarization in the signal is extremely unlikely to be reproduced by models in which QED vacuum
polarization effects are not accounted for.

In this paper there are presented the results of calculations of other vacuum birefringence
manifestation in high magnetic field, which could be studied only by polarization measurements
of pulsed hard emission.

Observations just in hard X and gamma rays are preferable to avoid
the mentioned above polarization
effects caused by the pulsar magnetosphere plasma influence, which
are more weak for high energy photons.

General covariance equations of non-linear electrodynamics in
post-Maxwellian approximation  ($|{\bf E}|>B_q,\
|{\bf B}|>B_q$)  have the form:
$${1\over \sqrt{-g}}{\partial \over \partial x^n}
\Big\{\sqrt{-g}Q^{mn}\Big\}=-{4\pi \over c}j^m, \eqno(1)$$
$${\partial F_{mn} \over \partial x^k}+ {\partial F_{nk}
\over \partial x^m}+{\partial F_{km} \over \partial x^n}=0,$$
where 
$$ Q^{mn}=\Big[1+\xi\big(\eta_1-2\eta_2\big)J_2\big)\Big]F^{mn}
+4\xi\eta_2F^{mi}F_{ik}F^{kn}$$
here  notations were introduced, such as determinant of the
metric tensor
$g$ and  $\xi =1/B^2_q.$

The magnitude of the dimensionless post-Maxwellian parameters
$\eta_1 $ and $ \eta_2$
depends on the choice of vacuum non-linear electrodynamic model.
In the Heisenberg-Euler non-linear
electrodynamics 
\cite{c23}, 
which
is direct consequence of quantum electrodynamics,
parameters  $\eta_1$ and $ \eta_2$ have the values
$\eta_1=\alpha /(45 \pi )=5.1\cdot 10^{-5}, \ \eta_2=7\alpha
/(180 \pi )= 9.0\cdot 10^{-5};$  in the Born-Infeld theory 
\cite{c24} 
they are equal
$\eta_1=\eta_2=a^2B_q^2/4,$ where to the constant $a^2$
 only lower estimation is known  $a^2> 1.2\cdot
10^{-32}\ G^{-2}.$ Effects of vacuum polarization in
strong electromagnetic fields were studied in 
\cite{c25}, 
birefringence and photon splitting for
uniform magnetic field were analyzed in 
\cite{c26}. 

Birefringence in vacuum and its influence on the emission spectrum and
photon propagation in the
neutron star magnetosphere were discussed in details in
\cite{c27}, 
where it was presented that vacuum
effects determine polarization properties of normal modes of
 electromagnetic waves propagating in the vicinity
of neutron star. It was also shown, that magnetic vacuum effects
can change the emission spectral parameters
that should lead to peculiarities in energy spectra of X and gamma
ray pulsars 
\cite[]{c28,c29}. 

Non-linear electrodynamic effects caused by non-uniform and
non-stationary magnetic field of a neutron
star including beam bending in the dipole magnetic field,
electron-positron pair production, photon splitting
and modulation by the low-frequency electromagnetic wave scattering
in magnetic field of inclined rotator are
discussed in 
\cite{c30}. 
Quantum
electrodynamic effect in accreting neutron stars, in
particular, one and two photon Compton scattering in the strong
magnetic field and its acting on the emitting
processes 
\cite[]{c31,c32,c33} 
and vacuum polarization effects
in the field of charged compact object were
also studied 
\cite{c34}. 

Calculations and modeling of non-linear electrodynamic influence of
pulsar magnetic field on the electromagnetic pulse polarization were
made repeatedly in many particular cases
and with the use of different coordinate systems
\cite[]{c35,c36,c37,c38}. 
However, in these cases coordinates were chosen from the terms of
convenience and simplicity of  theoretical calculations. Often it was
not appropriate from the point of
observation interpretation. In this work we avoid this drawback and
make calculations of pulsar magnetic field
impact on electromagnetic pulse polarization in such a way to make it
easier to interpret these effects in
space experiments.

It is necessary to note, that pulsar also have a strong gravitational
field, which is also
can impact on electromagnetic emission pulses. Because gravitational
field cans only bend the trajectories of
electromagnetic beams 
\cite[]{c39,c40}, 
and it is not change their
polarization, we will neglect gravitational effects in this work.

\section{ The law of hard emission pulse
propagation in the pulsar magnetic field according to the vacuum non-linear electrodynamics}

Let us consider a pulsar with radius $R_N,$ which magnetic dipole moment $\bf M$ rigidly connected with pulsar
"body". Then we suppose that the initial point of Cartesian system is placed into the pulsar center. Let us
directed mutually orthogonal axes $X,\ Y$ and $Z$ in such a way, that $Z$ axis will pass through the center of
the Earth, and directions of  $X,\ Y$ will be cleared later. Let us consider, that at some time $t_0$ the
short X or gamma-ray burst with arbitrary polarization was occur in the point ${\bf r}_s=\{x_s,y_s,z_s\}$ of
pulsar magnetosphere. This emission will propagate as pulses along different beams in different directions
underwent non-linear electrodynamic impact from pulsar magnetic field. Let us consider that detector on a
satellite is placed in the point ${\bf r}_d=\{x_d,y_d,z_d\}$  in the vicinity of the Earth, that allows
measurements of hard emission pulse polarization.

We will calculate here the polarization state of pulse coming to detector. To study process, which is
interesting for us, there is no any necessity to solve the system of non-linear differential equations in
partial derivatives of the first order (1).
As it was shown in \cite{c38}, 
propagation of electromagnetic pulse in outer persistent electromagnetic field according to vacuum non-linear
electrodynamics can be describes by equations for isotropic geodesic in an effective space-time with metric
tensor $G^{(1,2)}_{pm}$:
$${dk^i\over d\sigma}
+G_{(1,2)}^{im}\Big[ {\partial G^{(1,2)}_{mn}\over\partial  x^p} -{1\over 2}{\partial
G^{(1,2)}_{np}\over\partial  x^m}\Big] k^nk^p=0,\eqno(2)$$ where $\sigma$ is affine parameter, $x^n=\{x^0=ct,\
x^1=x,\ x^2=y,\ x^3=z\},$ $k^n=dx^n/d\sigma$ is isotropic four-dimensional wave vector:
$$G^{(1,2)}_{mi}{dx^m\over d\sigma}{dx^i\over d\sigma}=0.\eqno(3)$$
The metric tensor $G^{(1,2)}_{pm}$ in post Maxwellian approximation of  vacuum non-linear electrodynamics (1)
depends on the metric tensor $g_{im}$ of pseudo Riemannian space-time, which is the solution of Einstein
equations from the outer electromagnetic field tensor $F_{ip}$ and from the polarization of considering
electromagnetic wave:
$$G_{(1,2)}^{im}=g^{mi}+4\eta_{1,2}\xi F^{in}F_{n\cdot}^{\cdot m}, \eqno(4)$$
$$G^{(1,2)}_{mn}=g_{mn}-4\eta_{1,2}\xi F_{mk}F^{k\cdot}_{\cdot n},$$
where index "1" refers to the first linearly polarized normal mode, and  index "2" is refers to the second
normal mode, which has the orthogonal polarization relatively to the first one. The indexes of electromagnetic
field tensor $F_{mk}$ in the relation (4) can be upped with the use of metric tensor of pseudo Riemannian
space-time $g^{mi}.$ The tensor components $G_{(1,2)}^{im}$ and $G^{(1,2)}_{mn}$ with post Maxwellian accuracy
are inversed to each other:
$$G_{(1,2)}^{im}G^{(1,2)}_{mn}=\delta^i_n
+O(\xi^2F^{ip}F_{p m}F^{mk}F_{kn}),$$ where $\delta^i_n$  is the Kronecker  tensor.

Existence of two relations $G^{(1)}_{mn}$ and $G^{(2)}_{mn}$ or effective space-time metric tensor  means that
according to the vacuum non-linear  electrodynamics electromagnetic wave birefringence occurs, i.e.  at
$\eta_1\neq\eta_2$  any electromagnetic wave splits onto the two normal modes with mutually orthogonal linear
polarization, which pass in the outer field with non-equal velocities. As it was shown in
\cite{c41} 
occurrence of effective space-time with metric tensor $G^{(1,2)}_{pm}$,  does not mean that Einstein Principe
of equivalence violets in the electrodynamics. Presence of second terms in expression (4) means that not only
gravitational field described by the metric tensor $g_{nk},$  affects on the electromagnetic wave propagation,
but outer electromagnetic field also.

Equations (2)-(3) easier to solve, choosing as independent variable coordinate $z$, but not the affine
parameter $\sigma$.  In this case we have from the equations (2):
$${d^2 x^0\over
dz^2}=\big\{G^{3m}_{(1,2)}{d x^0\over dz}-G^{om}_{(1,2)}\big\} \big\{{\partial G_{nm}^{(1,2)}\over \partial
x^p}-{1\over 2} {\partial G_{np}^{(1,2)}\over \partial x^m}\big\}
 {d x^n\over dz}{d x^p\over dz},\eqno(5)$$
$${d^2 x\over dz^2}=\big\{G^{3m}_{(1,2)}{d x\over dz}-G^{1m}_{(1,2)}\big\} \big\{{\partial G_{nm}^{(1,2)}\over
\partial x^p}-{1\over 2} {\partial G_{np}^{(1,2)}\over \partial x^m}\big\}{d x^n\over dz}{d x^p\over dz},$$
$${d^2 y\over dz^2}=\big\{G^{3m}_{(1,2)}{d y\over dz}-G^{2m}_{(1,2)}\big\}
\big\{{\partial G_{nm}^{(1,2)}\over \partial x^p}-{1\over 2} {\partial G_{np}^{(1,2)}\over \partial x^m}\big\}
{d x^n\over dz}{d x^p\over dz}.$$ Equation (3)in this case has the form:
$$G^{(1,2)}_{mn}{dx^m\over dz}{dx^n\over dz}=0.\eqno(6)$$

Because the relation (6)is the first integral of equation (5), the equation system (5)-(6) is linearly
dependent. Thus, one of the equations of this system can be omitted. It is easier to choose the first equation
of system (5)  as such one. These equations allow us to find the equation for rays $x=x(z),\ y=y(z)$ and the
law of electromagnetic pulse propagation along these rays $x^0=ct=x^0(z).$ Because for each normal mode  we
are interested in only one ray  along which pulse passes from the point  ${\bf r}={\bf r}_s$ in the point
${\bf r}={\bf r}_d$, let us â® require that relations
$$x^0(z_s)=ct_0,\ \ x(z_s)=x_s, \eqno(7)$$
$$ y(z_s)=y_s,\ \  x(z_d)=x_d,\ \
y(z_d)=y_d.$$ should valid as the initial conditions to equations (5)-(6) . To solve equations (5) - (6) with
post-Maxwellian accuracy it is enough to take in Maxwellian approximation the pulsar dipole magnetic field
induction vector $\bf B$ components:
$${\bf B}={3({\bf M\ r}){\bf r}-{\bf M}r^2\over r^5}.$$
It means that non-zero components of pulsar magnetic field tensor  in relation (4) are:
$$F_{21}=-F_{12}=B_z, F_{13}=-F_{31}=B_y, F_{32}=-F_{23}=B_x.\eqno(8)$$
The non-zero  components of pseudo-Riemannian space-time metric tensor $g_{im}$ if to neglect the
gravitational field take a form:
$$g_{00}=1,\ g_{11}=g_{22}=g_{33}=-1.\eqno(9)$$
If to substitute relations (8) - (9) into expressions (4), we find the explicit form of non-zero components of
effective space-time metric tensor $G^{(1,2)}_{pm}$:
$$G^{(1,2)}_{00}=1,\eqno(10)$$
$$G^{(1,2)}_{\alpha\beta}=-\delta_{\alpha\beta}\Big\{1+4\xi\eta_{1,2}
\Big[{3({\bf M\ r})^2\over r^8}+{{\bf M}^2\over r^6}\Big]\Big\}+$$
$$+4\xi\eta_{1,2}\Big[{9({\bf M\ r})^2\over r^{10}}x_\alpha x_\beta
+{M_\alpha M_\beta\over r^6} -{3({\bf M\ r})\over r^8}\Big(M_\alpha x_\beta+x_\alpha M_\beta\Big)  \Big], $$
where $\alpha,\ \beta... =1,2,3,$ $x_\alpha=\{-x,-y,-z\},$  $M_\alpha=\{-M_x,-M_y,-M_z\}.$

The detailed solution of equations (5)-(6) with effective space-time metric tensor which are satisfied the
initial conditions (7) is presented in Attachment A. 
 Let us
write the law of electromagnetic pulse passing $x^0=x^0(z)$ along the ray connecting the points ${\bf r}_s$
and ${\bf r}_d:$
$$t_{1,2}(z)=t_0+{(z-z_s)\over c}+\eta_{1,2}\xi  \big[\
\widetilde{t}(z)-\widetilde{t}(z_s)\big],$$ where notation is introduced
$$\widetilde{t}(z)=\big[25(M_xx_s+M_yy_s)^2+q^2(16{\bf
M}^2-M_z^2)\big]\Big[ {z(3\rho^2+2q^2)\over  64\rho^4q^6}+ {3\over  64q^7}\hbox{atan}\Big( {z\over
q}\Big)\Big]+\eqno(A13)$$
$$ +{1\over  \rho^6q^2}\big[5z(M_xx_s+M_yy_s)^2
+16q^2M_z(M_xx_s+M_yy_s)+3M_z^2q^2z\big]+ {9\over 4\rho^8}\big[z(M_xx_s+M_yy_s)^2 -$$
$$-q^2M_z(2M_xx_s+2M_yy_s+M_zz)\big].$$
To study pulsar magnetic field impact on  hard emission pulse polarization let us assume that its pulse at the
point of its appearance had arbitrary polarization or was not polarized.
  Due to the vacuum non-linear
electrodynamics birefringence this pulse splits on two pulses with mutually orthogonal linear polarizations.
These pulses propagate in the pulsar magnetic field with non-equal velocities. They have fronts, which
coincided at the initial time. The front of faster mode comes to the detector before, than the front of slow
mode on a time interval equal $\Delta t=|t_2(z_d)-t_1(z_d)|.$  It means that during the time $\Delta t$ only
fast normal mode of a pulse will pass though the detector and it will detect the linear polarization at this
part of pulse.

After a time $\Delta t$ the front of other normal pulse mode comes to the detector. Superposition of these
normal modes in time produces in the detector emission with arbitrary polarization. Hence, according to the
vacuum non-linear electrodynamics the front
 part of any hard emission pulse with
duration $á\Delta t$) coming from a pulsar  should be linearly polarized and the rest of pulse can have
arbitrary polarization. Electromagnetic pulse of faster mode will pass though the detector before the pulse of
slow mode. After that the trailing edge of this pulse passed through the detector during a time $\Delta t$
only the trailing edge of slow mode pulse with orthogonal polarization will be detected. Thus, the "tail" with
duration $á\Delta t$ of any hard emission pulse underwent the non-linear
 electrodynamics impact of pulsar
magnetic field should be linearly polarized, and its polarization will be orthogonal to polarization of the
pulse front.

Thus, to test the predictions of vacuum non-linear electrodynamics, it is necessary control the polarization
state of hard emission pulses coming from magnetars and pulsars on all their duration. If to use the
expressions (A3) and (A13), we can calculate the time $\Delta t=|\Delta\tau|:$
$$\Delta \tau=t_2(z_d)-t_1(z_d)=(\eta_2-\eta_1)\xi\Big\{\Big[{3\pi\over128q^7}
-{z_s(3\rho^2+2q^2)\over  64r_s^4q^6}-{3\over  64q^7}\hbox{atan}\Big( {z_s\over  q}\Big)\Big]\times\eqno(11)$$
$$\times\big[25(M_xx_s+M_yy_s)^2+q^2(16{\bf M}^2-M_z^2)\big]-{1\over 8r_s^6q^2}\big[5z_s(M_xx_s+M_yy_s)^2+$$
$$+16q^2M_z(M_xx_s+M_yy_s)+3M_z^2q^2z_s\big]-
{9\over 4r_s^8}\big[z_s(M_xx_s+M_yy_s)^2
-q^2M_z(2M_xx_s+2M_yy_s+M_zz_s)\big],$$

where  $r_s=\sqrt{x_s^2+y_s^2+z_s^2}.$

\section{ Observational possibilities of vacuum
nonlinear electrodynamics effects}

The time interval $\Delta t$, which characterize delay of signals carried by electromagnetic waves polarized
normally to each other essentially depends on the difference of post-Maxwellian parameters $\eta_1-\eta_2$.
Thus, it is different in various models of vacuum nonlinear electrodynamics. In particular, in the
Heisenberg-Euler  electrodynamics
 this delay time may be about 1
mcs for typical pulsar, while in the Born-Infeld  theory it is strictly
 equal to zero.

 From the
experimental point of view the effect of delaying of electromagnetic
 signals emitted from the poles of
rotating neutron star will be revealed differently in the case of slow
 varying and burst-like or pulsed
emission. In the case of slow varying emission the time dependence of detected intensity on one polarization
mode will be shifted relatively to the time dependence of orthogonal mode intensity. This delay time will
depend on the angle $\beta$ between neutron star magnetic momentum vector and the radius-vector of the point
of detector place. Lag effect will manifest itself in another way in the case of burst-like or pulsed emission
(photon beam) which duration is higher than time $\Delta t$, defined
 by expression (11).

 In this case, if
arbitrary polarized pulses are emitted, the detected pulses will have
 the variable polarization along their
length, i.e. at $\eta_1 > \eta_2$ the pulse front part of duration
 $\tau = \Delta t$ should be polarized
normally to the neutron star magnetic meridian plane and the other part
 of a pulse will be polarized randomly
in general case. If we assume that neutron star angular velocity $\Omega$ is sufficiently small, then the
linear velocity of points on the neutron star surface will be much
 less than light velocity in vacuum, and
electromagnetic signal propagation time in the area $r \sim 5R_s$ of
 strong magnetic field, where the
non-linear electrodynamic and gravitational actions on these signals
 are most significant, will be much less
than the star rotation period. Thus, the area of non-linear
 electromagnetic action will be in the near field
of magneto-dipole emission, it was shown
\cite{c42} 
that the lag time of signals with two main polarizations will be modulated due to the star rotation. It is
 well-known that time profiles of pulsar X- and
gamma ray emission look like a sequence of pulses with period equal
 to the neutron star rotation period and
pulse duration is determined mainly by emission beam width.

In the
 case of very narrow beam the pulsar
emission time profile can be presented as periodic pulses, which
 duration much less than period. At $\eta_1 >
\eta_2$ obviously, that pulse front will be polarized linearly,
 which plane is normal to the neutron star
magnetic meridian. As it following from (11) duration of this
 polarized pulse front will change from zero to
maximal value according to (11). Because the pulse duration $\tau$ is less than period $2\pi/\Omega$, periodic
change of intensity for time $\tau$ will not significant, thus for a real analysis it is possible to assume
that polarized part of a pulse is constant. Its value will depend on observation line direction relatively to
the neutron star meridian plane.

It was noted above that vacuum non-linear electrodynamic effects caused by the acting of very strong fields
(including birefringence) are the same as non-linear effects in a matter. In this case strong magnetic field
can be considered as effective matter. Thus, emission parameters determined by vacuum non-linear
electrodynamic effects can be masked principally by non-linear influence of a matter. In particular, due to
the Faraday effect, non-polarized emission scattered on plasma electrons becomes partially linearly polarized
due to the inhomogeneous distribution of polarization plane rotation angles.

Such effect was analyzed for the cases of light propagation in magnetized stellar wind
\cite{c43}, 
strongly magnetized optically thick
accretion disk 
\cite{c44}, 
magnetized cone shell, which can be considered as model of relativistic
 jet 
 \cite{c45}. 
 However,
non-linear electrodynamic effects caused by the matter influence will
 dominate on vacuum effects mainly for
photons from long wave bands of electromagnetic spectrum, i.e. optics
 and radio. Indeed, polarization
properties of matter are determined significantly by dielectric constant $\epsilon$, for which the expression
\cite{c46} 
is well-known:
$$ \varepsilon=1-{4\pi N_e\hbar^2e^2\over m_eE_\gamma^2}, \eqno(12)$$
where $E_\gamma$ is the photon energy, $N_e$ is the electron concentration in plasma, $m_e$ is the electron
mass, h is the Plank constant. If to substitute in eq. (12) the typical $N_e \sim 10^{19}$ cm$^3$ then we
obtain for $E_\gamma = 0.1$ MeV $\varepsilon = 1 - 10^{-11}$.

Thus, the matter influence on polarization properties of X ray and gamma emission is pitifully. This
conclusion is confirmed by digital calculations, which shows that polarization degree of emission after it
passed through magnetic plasma decreases with decreasing of its wave
length 
\cite{c47}. 

Vacuum non-linear effects also can be connected with some exotic processes, for example, with mutual
transformation in magnetic field of photons and light Goldstone bosons (axions). Electromagnetic emission
occurred from the axion decay in magnetic
field can be highly polarized 
\cite{c48}. 
However, according to
\cite{c48} 
axion input in dielectric constant is inversely proportional to the square of the photon frequency (or energy)
(compare with (12)). Thus, such processes can't be significant for determination of polarization properties of
high energy photons.

So, it could be expected that revealing of vacuum non-linear electrodynamic effects will be most pure just in
X and gamma rays from pulsars. It was shown above, that non-linear electrodynamics birefringence of X and
gamma rays in strong magnetic field of pulsars and magnetars can be revealed as about 1 mcs lag of two signal
modes almost $100\%$ polarized in mutually normal directions. To measure such lag experimentally, it is
necessary to realize accurate polarization measurements allowing obtain the mean pulsation curve for linearly
polarized and non-polarized pulse components with fine time resolution (about 0.1 mcs). It is also necessary
to measure with high accuracy ($\sim 0.1$ mcs) the arrival time of detected quanta
 during the all-time of observations.

 To make
a final choice in favor of one or another experimental method, let us
 analyze the main factors that determine
sensitivity of polarization measurements. For example, consider the
 Compton scattering technique as the most
universal method of polarization measurements in hard X ray and gamma
 ray astronomy. Let $I$ is the total
number of signal counts, $N$ is the background (noise) counts. Then for
 expected signal to noise ratio
expressed in number of standard deviation $\sigma$ for given exposure
 time we have for polarization
measurements an expression:
$$\sigma={\mu IP\over \sqrt{2(I+N)}},\eqno(13)$$
where $P$ is the polarization degree and $\mu$ is the instrumental
 polarization factor, in which source
position on the sky is also taken into account.

From (13) it is possible
 to obtain the estimation of minimally
detectable polarization degree $P_{min}$.

    As it following from (13), to increase sensitivity it is necessary
 to minimize nose, increase the useful signal
and increase the instrumental polarization factor $\mu$. The number of
 useful signal counts depends
on the detected flux intensity $J$, which is expressed in cm$^{-2}s^{-1}$ and equal to
$$I = JS_{(eff)} \Delta t,    \eqno(14)$$
where $S_{eff}$ is the effective detector area, $\delta t$ is the time set of signal statistics, which in the
case of burst or pulse like signal is equal to the burst (pulse) duration and in the case of slow varying
signals is determined
 by the exposure time.

 Thus, it is
following from (14) that to increase the useful signal it is necessary
 maximally increase the detector
effective area and the exposure time. Thus, the increasing of detector
 area is inevitably linked with
increasing of its mass and sizes, it is necessary to estimate the
 reasonable limits, which are defined of
course by the source observable luminosity.

To estimate conditions for detection of $100\%$ polarization, i.e. $P_{min}$=1, let us put $\mu$=1 and neglect
noise. Then to provide the 3$\sigma$ significance level no less than about 25 counts should be detected at
each time interval (bin), on which polarization is measured. To reveal the lag
 in about 1 mcs time bin should be not more
than 1 mcs also at least.

Then in the case of continuous signal from
 pulsar no less than $25T/(2(1\  mcs))$
counts should be accumulated for pulsation period $T$. For simplicity
 we assume that pulsation profile is as
rectangular with pulse phase duration equal to the pulsation period $T$.
 If the intensity of signal pulsed
component is $J$ (in cm-2s-1), then in the case of detector with
 effective area $S_{eff}$ for one pulsation
period $J\odot T\odot S_{eff}/2$ counts will be detected. Then to
 satisfy the condition of $100\%$ polarization
detection with 1 mcs time resolution the following expression should
 be valid:
$${25T\over(1\ mcs)} = J S_{eff} T n,\eqno(15)$$
where $n$ is the number of pulsation periods for a total exposure time.

Equality (15) can be rewritten as
$${25T\over(1\ mcs)} = J S_{eff} \Delta t,$$
 where $\Delta t$ is the total exposure time of a source.

It is well-known, that Crab pulsar is the most intensive in different
 energy
bands among the other pulsars. Its spectrum can be good approximated
 in wide energy range by power law with
power index $\sim 2$ 
\cite[]{c49,c50}, 
Then for the Crab pulsar intensity the following estimations can be obtained for different energies: $J (E  =
20-100\ keV) = ~4.6 10^{-2}$ phot/cm$^2$/s, $J (E= 0.1-1.0\ MeV) = ~1.2 10^{-2}$ phot/cm$^2$/s. Taking into
account
 that Crab pulsation period is equal 33 ms, we
obtain from (15) the estimations of factor $S_{eff}\Delta t$.

For energy
 range 20 - 100 keV effect can be
detected in the case of 1 Crab intensity, if $S_{eff}$ = $10^3$ $cm^2$,
$\Delta t$ $\sim$ 100 ks and for 1 mCrab
intensity, if $S_{eff}$ = $10^4$ $cm^2$, $\Delta t\sim $ 3 Ms. For energy
 range 0.1 - 1.0 MeV we have
respectively if $S_{eff}$ = $10^4$ $cm^2$, $\Delta t\sim$ 100 ks for 1 crab and if $S_{eff} = 10^4$ $cm^2$,
$\Delta t$ $\sim$ 10 Ms for 1 mCrab It is necessary to note, that effective area is determined not only by
detector geometry area, but also by efficiency of scattering or any
 other process, which is used for
polarization measurements. For the all well-known polarization techniques, including Compton polarimeters,
efficiency no more than 10$\%$, really it is about a few percent. Thus,
 the polarimeter geometry area should
be taken at least on order more than obtained above estimations of its
 effective area.

\begin{table}
    \centering
    \caption{Signal to noise ratio of polarized fraction of pulsar emission in different energy
    range for different background models.}
    \label{tab1}
    \begin{tabular}{lcccr} 
        \hline
         & Energy & ranges,& MeV& \\
        \hline
        SNR & 0.002 - 0.001 & 0.05 - 0.1 & 0.5 - 1.0& 100-500\\
\hline
        ${I\over N_{diff}}$ & 1 & 6 & 150& 250\\
\hline
        ${I\over (N_{diff}+N_{int})}$ & 1 & 5 & 6&0.25\\
        \hline
    \end{tabular}
\end{table}


To choose the optimal observational method, it is necessary to analyze
 the main background factors in
different energy ranges. Generally, detector background counts $N$ can
 be presented as a sum of parts caused
by natural or Galactic and Meta-galactic diffuse background $N_{diff}$
 and intrinsic background of detector and
satellite $N_{int}$. Possible signal to noise ratio (SNR) can be
 estimated from pulsar energy spectrum, which
can be taken typically as Crab-like. The corresponding SNR values are
 presented in the Table 1 in relative
units, in which SNR in the 2 - 20 keV range is taken equal to 1. These
 values are obtained for a instrument
with FOV $2\pi$ sr, i.e. intensities of all background components were
 multiplied on $2\pi$.
 \
 As it follows
from the Table 1, if take into account only the natural background, the best SNR value is obtained for the
ranges of soft (0.1-1.0 MeV) and high energy (0.1-0.5 GeV) gamma rays.
 However, it is necessary to note, that
at the energies less than 0.1 MeV Meta-Galactic diffuse background is
 dominating and its input in detector
counts is proportional to the instrument FOV. While at energies >0.1 MeV
 noise counts are determined mainly by
spacecraft and detector intrinsic background and weakly depends on FOV.

 The SNR values, in which contribution
of intrinsic background was taken into account, are also presented in
 the Table 1. In this case, SNR has
favorable values for the range of hard X rays and soft gamma rays. It
 can be explained by that energy spectrum
of intrinsic background is extremely hard, (see, for
ex. 
\cite{c51} 
) in presentation $E^J$ it can even grow with energy. Of course intrinsic background depends strongly
 on the spacecraft mass, which should be
as low as possible, but such energy dependence of intrinsic background
 intensity is valid for satellites.

\section{Conclusions}

Thus, we can conclude, that the optimal energy range for vacuum
 non-linear electrodynamics effect search is
the range 0.05 - 1.0 MeV, which can be extended up to about 10 MeV.
 What about the factor $S_{eff}\Delta t$, it
should be understood, that increasing of the instrument geometry area
 can't be infinite due to the limited
resources of space experiments. Obviously, that detector area about
 $10^4 cm^2$ is near that limit, which can
be still realized in space experiment. Concerning the exposure time,
 it can be made about of all time of
experiment, which can be about of years, in the case of constant
 orientation of the instrument on a source or
in the case of monitor observations with wide FOV telescope.

Thus,
 just the long time monitor observations are
necessary for the search and revealing of non-linear electrodynamics effects. In view of discussed above
features it seems that use of monitor instruments based on the Compton
 polarimeters are the most realistic way
to realize the vacuum non-linear electrodynamics birefringence search
 and observation. Compton polarimeters
have a real advantage in view of SNR optimization, because they based on detection of pair coincidences of
incident and scattered photons, that allows eliminate instrument background effectively. Besides, such
instruments give an opportunity to locate the source of detected gamma
 quanta by revealing of useful signals
from the background cause by spacecraft and detector intrinsic noise
 and the atmosphere gamma rays (in the
case of near-Earth observations). By this, the most dangerous are
 random coincidences in limits of trigger
time $\tau$ of those noise events, which can be detected independently
 in the diffuse and detector of
scattered quanta. As it is well-known, if noise count $N$ is constant
 and equal in diffuser and detector, the
count of random coincidences $M$ is equal $M = N^\tau$. Thus, value $M$
 can be made negligible in the case of
very small time window $\tau$, that is in accordance with discussed
 above time resolution about 0.1 mcs.

Because in the Compton process quanta scatter presumably at 90$^o$ relative to the incidence direction and in
the case of linear polarization they are scattered normally to the
 polarization plane, intensity of scattered
quanta will be modulated harmonically on Azimuth angle. Measurements
 by Compton polarimeters mean the
obtaining of histogram of coincidence pair distribution on Azimuth
 angle and its approximation by harmonic
function. Then this function amplitude can be used as the measure of
 polarization degree. The main factor
worsening the instrument polarization capabilities is Coulomb
 scattering, which decreases exponentially the
instrument polarization factor and minimal detectable polarization
 degree respectively. For example, CGRO
EGRET instrument intended for observations in the range of high energy gamma rays scatter factor decrease
effectively polarization factor in $10^{-4}$ times that made this
 experiment non-sensitive to polarization
measurements.

To increase the efficiency of polarimetry measurements
 it is necessary to maximize the number of
pair coincidences corresponding to interactions in two neighbor
 detector pixels. For this the radius of
detector pixel should be about one mass absorption length $\lambda$.
 In the case of most popular for gamma
quantum detection dense crystals as semiconductor as CdZnTe for example,
 or scintillator as $LaBr_3$, $CeBr_3$
and Ce:GAGG the pixel radius should be choose about 0.5-1.0 cm. Thus,
 the Compton polarimeter instrument which
is useful for vacuum non-linear electrodynamics birefringence
 observations should be based on large area
($\sim$ $10^4$ $cm^2$) detector consists from small pixels of about
 1 cm size.

 It should be taken into
account, that  in most Compton telescopes the reconstruction of the
 direction of the incident photon provides
an uncertainty area which has the shape of a thin cone arc. However,
 there is an alternative way to use a gas
TPC, that is also called Electron Tracking Compton Camera (ETCC), which
 provides the tracking of the recoil
electron from the first Compton interaction with a measurement of the
 direction of the recoil momentum. It
allows to decrease the length of the arc and therefore to improve
 dramatically the sensitivity of the detector
\cite{c21}, 
and references therein).

Some of these
 telescopes are sensitive to photon energies up
to tens of MeV in the Compton mode, but their sensitivity to
 polarisation above a few MeV is either
nonexistent or undocumented. Wide-field gamma ray (0.02 - 3.0 MeV)
 telescope Gammascope, which is elaborated
now at SINP MSU in the frame of Russian space program is able to realize the polarization measurements based
on Compton technique. This instrument is the position sensitive detector
 (PSD) with coding mask of
quasi-spherical (dodecahedron) configuration. It FOV of about $2\pi$
 sr provides the continuous observation of
a half of the sky during space experiment.

The instrument should consist
 from six PSD modules placed on the
bottom parts of dodecahedron frame and from six coding mask panels from
 Wf or Ta placed on the top parts of
dodecahedron frame, i.e. the each PSD module has an opposite coding mask
 panel and all system observes half of
sky. Due to such configuration the exposure time of each source in the
 instrument FOV will be about the
all-time of experiment, i.e. few years. It is supposed that PSD should
 consists from a large number of
cylindrical scintillator ($CeBr_3$ or Ce:GAGG) pixels of about 0.5 cm
 diameter and 2 cm height. The dense
scintillator like CsI(Tl) or BGO will be used as active shield. The total area in the compact configuration
should be $\sim$ $10^3$ $cm^2$ and $\sim$ $10^4$ $cm^2$ in the optimal
 case.

 The special mode of double
coincidences is foreseen for polarization measurements. The combination
 of large area and extremely long
exposure time gives the good opportunity to search the fine polarization
 effects like vacuum non-linear
electrodynamics birefringence.

\appendix
\section{Detailed calculations }

Let us start the solving of the equation system (5)-(6) with initial
 conditions (7). With the use of relation (10), we then obtain for
the first integral (6):
$$c^2\left({dt\over dz}\right)^2-\left({d{\bf r}\over dz}\right)^2
\Big\{1+4\xi\eta_{1,2} \Big[{3({\bf M\ r})^2\over r^8}+{{\bf M}^2\over r^6}\Big]\Big\}+\eqno(A1)$$
$$+{4\xi\eta_{1,2}\over r^6}\Big[{3({\bf M\ r})\over r^2}
\left({\bf r}\ {d{\bf r}\over dz}\right) -\left({\bf M}\ {d{\bf r}\over dz}\right)\Big]^2=0. $$
Substituting
(10) into the last two equations of system (5),  we obtain:
$${d^2x\over dz^2}=\eta_{1,2}\xi \Big\{\Big[
{180\over r^{12}}({\bf M\ r})^2x^2z+{12\over r^{10}}({\bf M\ r})
[9({\bf M\ r})z
-z(14xM_x+8yM_y)+\eqno(A2)$$
$$(3x^2+8y^2)M_z]+{12\over r^8}[z({\bf M}^2+M_z^2)
-9({\bf M\ r})M_z+(3xM_x+yM_y)M_z]\Big]
\left({dx\over dz}\right)^3+\Big[
{360\over r^{12}}({\bf M\ r})^2xyz-$$
$$-{24\over r^{10}}({\bf M\ r})[3z(y M_x+xM_y)+5xyM_z]+{24\over r^8}M_z[xM_y+yM_x]\Big]
\left({dx\over dz}\right)^2\left({dy\over dz}\right)+$$
$$+\Big[
{180\over r^{12}}({\bf M\ r})^2y^2z+{12\over r^{10}}({\bf M\ r})
[9({\bf M\ r})z+(8x^2+3y^2)M_z-z(8xM_x+14yM_y)]+$$
$$+{12\over r^8}[z({\bf M}^2+M_z^2)+(xM_x+3yM_y)M_z-9({\bf M\ r})M_z
]\Big]
\left({dx\over dz}\right)\left({dy\over dz}\right)^2+$$
$$+\Big[
{24\over r^{10}}({\bf M\ r})
[15({\bf M\ r})y+3(2x^2+y^2)M_y+8y(xM_x-zM_z)]
-{360\over r^{12}}({\bf M\ r})^2y(2x^2+$$
$$+y^2)+{24\over r^8}[(zM_z-xM_x)M_y-3({\bf M\ r})M_y+y(M_z^2-M_x^2)]\Big]
\left({dx\over dz}\right)\left({dy\over dz}\right)+$$
$$+\Big[
{12\over r^{10}}({\bf M\ r})[21({\bf M\ r})x+25x^2 M_x
+8x(yM_y-zM_z)+6y^2M_x]-$$
$$-{180\over r^{12}}({\bf M\ r})^2x(3x^2+2y^2)
-{12\over r^8}[5({\bf M\ r})M_x+x(4M_x^2+M_y^2-M_z^2)
+$$
$$+(yM_y-zM_z)M_x]\Big]\left({dx\over dz}\right)^2
+\Big[
{12\over r^{10}}({\bf M\ r})[(8x^2+5y^2)M_x-9({\bf M\ r})x+x(14yM_y+$$
$$+8zM_z)]
-{180\over r^{12}}({\bf M\ r})^2xy^2
+{12\over r^8}[({\bf M\ r})M_x-x(2M_x^2+M_y^2+M_z^2)
-$$
$$-(3yM_y+zM_z)M_x]\Big]\left({dy\over dz}\right)^2
+\Big[
{12\over r^{10}}({\bf M\ r})[24z({\bf M\ r})+(25x^2+19y^2) M_z+$$
$$+8z(xM_x-yM_y)]-{180\over r^{12}}({\bf M\ r})^2z(3x^2+y^2)
-{12\over r^8}[20({\bf M\ r})M_z+(xM_x-yM_y)M_z+$$
$$+z(M_x^2-M_y^2-4M_z^2)]
\Big]\left({dx\over dz}\right)
+\Big[
{24\over r^{10}}({\bf M\ r})[3xyM_z+3xzM_y+5yzM_x]-$$
$$-{360\over r^{12}}({\bf M\ r})^2xyz-{24\over r^8}
(yM_z+zM_y)M_x\Big]\left({dy\over dz}\right)+
{180\over r^{12}}({\bf M\ r})^2x(x^2+y^2)+$$
$$+{12\over r^{10}}({\bf M\ r})[(3x^2-5y^2) M_x-24({\bf M\ r})x
+x(8yM_y+14zM_z)]+$$
$$+{12\over r^8}[6({\bf M\ r})M_x-x(2M_x^2+M_y^2+M_z^2)
-(yM_y+3zM_z)M_x]\Big\},$$
$${d^2y\over dz^2}=\eta_{1,2}\xi \Big\{\Big[
{180\over r^{12}}({\bf M\ r})^2y^2z+{12\over r^{10}}({\bf M\ r})
[9({\bf M\ r})z-z(14yM_y+8xM_x)+(3y^2+8x^2)M_z]+$$
$$+{12\over r^8}[z({\bf M}^2+M_z^2)
-9({\bf M\ r})M_z+(3yM_y+xM_x)M_z]\Big]
\left({dy\over dz}\right)^3+\Big[
{360\over r^{12}}({\bf M\ r})^2xyz-$$
$$-{24\over r^{10}}({\bf M\ r})[3z(x M_y+yM_x)
+5xyM_z]+{24\over r^8}M_z[xM_y+yM_x]\Big]
\left({dy\over dz}\right)^2\left({dx\over dz}\right)+$$
$$+\Big[
{180\over r^{12}}({\bf M\ r})^2x^2z+{12\over r^{10}}({\bf M\ r})
[9({\bf M\ r})z+(8y^2+3x^2)M_z-z(8yM_y+14xM_x)]+$$
$$+{12\over r^8}[z({\bf M}^2+M_z^2)+(yM_y+3xM_x)M_z-9({\bf M\ r})M_z
]\Big]
\left({dy\over dz}\right)\left({dx\over dz}\right)^2+$$
$$+\Big[
{24\over r^{10}}({\bf M\ r})
[15({\bf M\ r})x+3(2y^2+x^2)M_x+8x(yM_y-zM_z)]
-{360\over r^{12}}({\bf M\ r})^2x(2y^2+$$
$$+x^2)+{24\over r^8}[(zM_z-yM_y)M_x-3({\bf M\ r})M_x+x(M_z^2-M_y^2)]\Big]
\left({dy\over dz}\right)\left({dx\over dz}\right)+$$
$$+\Big[
{12\over r^{10}}({\bf M\ r})[21({\bf M\ r})y+25y^2 M_y
+8y(xM_x-zM_z)+6x^2M_y]-$$
$$-{180\over r^{12}}({\bf M\ r})^2y(3y^2+2x^2)
-{12\over r^8}[5({\bf M\ r})M_y+y(4M_y^2+M_x^2-M_z^2)
+$$
$$+(xM_x-zM_z)M_y]\Big]\left({dy\over dz}\right)^2
+\Big[
{12\over r^{10}}({\bf M\ r})[(8y^2+5x^2)M_y-9({\bf M\ r})y+y(14xM_x+$$
$$+8zM_z)]
-{180\over r^{12}}({\bf M\ r})^2yx^2
+{12\over r^8}[({\bf M\ r})M_y-y(2M_y^2+M_x^2+M_z^2)
-$$
$$-(3xM_x+zM_z)M_y]\Big]\left({dx\over dz}\right)^2
+\Big[
{12\over r^{10}}({\bf M\ r})[24z({\bf M\ r})+(25y^2+19x^2) M_z+$$
$$+8z(yM_y-xM_x)]-{180\over r^{12}}({\bf M\ r})^2z(3y^2+x^2)
-{12\over r^8}[20({\bf M\ r})M_z+(yM_y-xM_x)M_z+$$
$$+z(M_y^2-M_x^2-4M_z^2)]
\Big]\left({dy\over dz}\right)
+\Big[
{24\over r^{10}}({\bf M\ r})[3xyM_z+3yzM_x+5xzM_y]-$$
$$-{360\over r^{12}}({\bf M\ r})^2xyz-{24\over r^8}
(xM_z+zM_x)M_y\Big]\left({dx\over dz}\right)+
{180\over r^{12}}({\bf M\ r})^2y(x^2+y^2)+$$
$$+{12\over r^{10}}({\bf M\ r})[(3y^2-5x^2) M_y-24({\bf M\ r})y
+y(8xM_x+14zM_z)]+$$
$$+{12\over r^8}[6({\bf M\ r})M_y-y(2M_y^2+M_x^2+M_z^2)
-(xM_x+3zM_z)M_y]\Big\}.$$

Equations (A1) and (A2) are non-linear, for which the conventional methods of
solution \cite{c54} are not applicable. However, they contain the small parameter $\xi {\bf M}^2/r^6.$
Therefore successive approximation is convenient method to solve these equations. For this, we substitute
$x=x_{1,2}(z),\ y=y_{1,2}(z)$ and $t=t_{1,2}(z)$ In the form of expansions on the small parameter:
$$t_{1,2}(z)=T(z)+\eta_{1,2}\xi
\big[\ \widetilde{t}(z)-\widetilde{t}(z_s)\big],\eqno(A3)$$
$$x_{1,2}(z)=X(z)+\eta_{1,2}\xi
\Big[\ \widetilde{x}(z)- \widetilde{x}(z_d)+{(z_d-z)\over (z_d-z_s)} \big[\
\widetilde{x}(z_d)-\widetilde{x}(z_s)\big]\Big],$$
$$y_{1,2}(z)=Y(z)+\eta_{1,2}\xi
\Big[\ \widetilde{y}(z)-\widetilde{y}(z_s)+{(z_d-z)\over (z_d-z_s)} \big[\
\widetilde{y}(z_d)-\widetilde{y}(z_s)\big]\Big],$$ where $T(z),X(z),Y(z)$ are unknown functions of zero
approximation,
  nd $\ \widetilde{t}(z), \widetilde{x}(z),\widetilde{y}(z)$ are of the first approximation.

It is following from expressions (7), that initial conditions for the functions in these expressions have the
form:
$$T(z_s)=ct_0,\ X(z_s)=x_s,\ Y(z_s)=y_s,\ X(z_d)=x_d,\ Y(z_d)=y_d.\eqno(A4)$$
In the Maxwellian approximation equations (A1)-(A2) give:
$$c^2{d^2T(z)\over dz^2}={d^2X(z)\over dz^2}={d^2Y(z)\over dz^2}=0,$$
$$c^2\left({dT(z)\over dz}\right)^2-\left({dX(z)\over dz}\right)^2
-\left({dY(z)\over dz}\right)^2=1.$$

It is easy to obtain the solution of these equations satisfying the initial conditions (€4):
$$T(z)=t_0+{N\over c}(z-z_s),\ {\bf R}={\bf r}_s+(z-z_s){\bf N},\eqno(A5)$$
where notations
$${\bf R}=\{X(z), Y(z), z\}, \ \ N=|{\bf N}|,\eqno(A6)$$
$$N_x={(x_d-x_s)\over(z_d-z_s)},\
N_y={(y_d-y_s)\over(z_d-z_s)},\ N_z=1.$$ are used for the convenience of further calculations. With the use of
expressions (A5) equations (A2) in the post-Maxvellian
 approximations have the form:
$${d^2\widetilde{x}(z)\over dz^2}={1\over N^{12}} \Big\{\Big[
{180\over \rho^{12}}({\bf M\ R})^2X^2z+{12N^2\over \rho^{10}}({\bf M\ R}) [9({\bf M\ R})z-z(14XM_x+8YM_y)+
 \eqno(A7)$$
$$+(3X^2+8Y^2)M_z]+{12N^4\over \rho^8}[z({\bf M}^2+M_z^2)-9({\bf M\ R})M_z+(3XM_x+YM_y)M_z]\Big]N_x^3+\Big[
{360\over \rho^{12}}({\bf M\ R})^2XYz+$$
$$+{24N^4\over \rho^8}M_z[XM_y+YM_x]-{24N^2\over \rho^{10}}({\bf M\ R})[3Yz M_x
+3XzM_y+5XYM_z]\Big]N_x^2N_y+\Big[
{180\over \rho^{12}}({\bf M\ R})^2Y^2z+$$
$$+{12N^2\over \rho^{10}}({\bf M\ R}) [9({\bf M\ R})z
+(8X^2+3Y^2)M_z-z(8XM_x+14YM_y)]+{12N^4\over \rho^8}[z({\bf M}^2+M_z^2)+$$
$$+(XM_x+3YM_y)M_z-9({\bf M\ R})M_z]\Big]N_xN_y^2+\Big[
{24N^2\over \rho^{10}}({\bf M\ R})[15({\bf M\ R})Y+3(2X^2+Y^2)M_y+$$
$$+8Y(XM_x-zM_z)]-{360\over \rho^{12}}({\bf M\ R})^2Y(2X^2+Y^2)
+{24N^4\over \rho^8}[(zM_z-XM_x)M_y-3({\bf M\ R})M_y+$$
$$+Y(M_z^2-M_x^2)]\Big]N_xN_y
+\Big[ {12N^2\over \rho^{10}}({\bf M\ R})[21({\bf M\ R})X+25X^2 M_x+6Y^2M_x+8X(YM_y-zM_z)]-$$
$$-{180\over \rho^{12}}({\bf M\ R})^2X(3X^2+2Y^2)-{12N^4\over \rho^8}[5({\bf M\ R})M_x+X(4M_x^2+M_y^2-M_z^2)+(YM_y-zM_z)M_x]\Big]N_x^2+$$
$$+\Big[
{12N^2\over \rho^{10}}({\bf M\ R})[(8X^2+5Y^2)M_x-9({\bf M\ R})X+X(14YM_y+8zM_z)]
-{180\over \rho^{12}}({\bf M\ R})^2XY^2+$$
$$+{12N^4\over \rho^8}[({\bf M\ R})M_x-X(2M_x^2+M_y^2+M_z^2)-(3YM_y+zM_z)M_x]\Big]N_y^2
+\Big[ {12N^2\over \rho^{10}}({\bf M\ R})[24z({\bf M\ R})+(25X^2+$$
$$+19Y^2) M_z
+8z(XM_x-YM_y)]-{180\over \rho^{12}}({\bf M\ R})^2z(3X^2+Y^2)-{12N^4\over \rho^8}[20({\bf M\ R})M_z+$$
$$+(XM_x-YM_y)M_z
+z(M_x^2-M_y^2-4M_z^2)]\Big]N_x+\Big[
{24N^2\over \rho^{10}}({\bf M\ R})[3XYM_z+3XzM_y+5YzM_x]-$$
$$-{360\over \rho^{12}}({\bf M\ R})^2XYz-{24N^4\over \rho^8}
(YM_z+zM_y)M_x\Big]N_y+ {180\over \rho^{12}}({\bf M\ R})^2X(X^2+Y^2)+$$
$$+{12N^2\over \rho^{10}}({\bf M\ R})[(3X^2-5Y^2) M_x-24({\bf M\ R})X
+X(8YM_y+14zM_z)]+$$
$$+{12N^4\over \rho^8}[6({\bf M\ R})M_x-X(2M_x^2+M_y^2+M_z^2)
-(YM_y+3zM_z)M_x]\Big\},$$

$${d^2\widetilde{y}(z)\over dz^2}={1\over N^{12}} \Big\{\Big[
{180\over \rho^{12}}({\bf M\ R})^2Y^2z+{12N^2\over \rho^{10}}({\bf M\ R}) [9({\bf M\ R})z
-z(14YM_y+8XM_x)+(3Y^2+8X^2)M_z]+$$
$$+{12N^4\over \rho^8}[z({\bf M}^2+M_z^2)-9({\bf M\ R})M_z+(3YM_y+XM_x)M_z]\Big]
N_y^3+\Big[ {360\over \rho^{12}}({\bf M\ R})^2XYz+$$
$$+{24N^4\over \rho^8}M_z[XM_y+YM_x]-{24N^2\over \rho^{10}}({\bf M\ R})[3z(X M_y+YM_x)+5XYM_z]\Big]
N_y^2N_x+$$
$$+\Big[ {180\over \rho^{12}}({\bf M\ R})^2X^2z+{12N^2\over \rho^{10}}({\bf M\ R})
[9({\bf M\ R})z+(8Y^2+3X^2)M_z-z(8YM_y+14XM_x)]+$$
$$+{12N^4\over \rho^8}[z({\bf M}^2+M_z^2)+(YM_y+3XM_x)M_z-9({\bf M\ R})M_z
]\Big]N_yN_x^2+\Big[
{24N^2\over \rho^{10}}({\bf M\ R}) [15({\bf M\ R})X+$$
$$+3(2Y^2+X^2)M_x+8X(YM_y-zM_z)]-{360\over \rho^{12}}({\bf M\ R})^2X(2Y^2+X^2)
+{24N^4\over \rho^8}[(zM_z-YM_y)M_x-$$
$$-3({\bf M\ R})M_x+X(M_z^2-M_y^2)]\Big]N_yN_x+\Big[
{12N^2\over \rho^{10}}({\bf M\ R})[21({\bf M\ R})Y+25Y^2 M_y +6X^2M_y+8Y(XM_x-$$
$$-zM_z)]-{180\over \rho^{12}}({\bf M\ R})^2Y(3Y^2+2X^2)
-{12N^4\over \rho^8}[5({\bf M\ R})M_y+Y(4M_y^2+M_x^2-M_z^2)+$$
$$+(XM_x-zM_z)M_y]\Big]N_y^2+\Big[
{12N^2\over \rho^{10}}({\bf M\ R})[(8Y^2+5X^2)M_y-9({\bf M\ R})Y+Y(14XM_x+$$
$$+8zM_z)]
-{180\over \rho^{12}}({\bf M\ R})^2YX^2 +{12N^4\over \rho^8}[({\bf M\ R})M_y
-Y(2M_y^2+M_x^2+M_z^2)-(3XM_x+zM_z)M_y]\Big]N_x^2+$$
$$+\Big[
{12N^2\over \rho^{10}}({\bf M\ R})[24z({\bf M\ R})+(25Y^2+19X^2) M_z+8z(YM_y-XM_x)]
-{180\over \rho^{12}}({\bf M\ R})^2z(3Y^2+X^2)-$$
$$-{12N^4\over \rho^8}[20({\bf M\ R})M_z+(YM_y-XM_x)M_z
+z(M_y^2-M_x^2-4M_z^2)]\Big]N_y+\Big[ {24N^2\over \rho^{10}}({\bf M\ R})\times$$
$$\times[3XYM_z+3YzM_x+5XzM_y]
-{360\over \rho^{12}}({\bf M\ R})^2XYz-{24N^4\over \rho^8}(XM_z+zM_x)M_y\Big]N_x+$$
$$+{180\over \rho^{12}}({\bf M\ R})^2Y(X^2+Y^2)
+{12N^2\over \rho^{10}}({\bf M\ R})[(3Y^2-5X^2) M_y-24({\bf M\ R})Y
+Y(8XM_x+14zM_z)]+$$
$$+{12N^4\over \rho^8}[6({\bf M\ R})M_y-Y(2M_y^2+M_x^2+M_z^2)
-(XM_x+3zM_z)M_y]\Big\},$$ where
$$\rho=\sqrt{(z+p)^2+q^2},\ \ X=x_s+N_x(z-z_s),\
Y=y_s+N_y(z-z_s),\eqno(A8)$$
$$p={[(x_d-x_s)(x_sz_d-z_sx_d) +(y_d-y_s)(y_sz_d-z_sy_d)]\over
({\bf r}_d-{\bf r_s})^2},$$
$$q^2={(z_d-z_s)^2\over({\bf r}_d-{\bf r_s})^4}\big[
(x_sy_d-y_sx_d)^2+(x_sz_d-z_sx_d)^2+(y_sz_d-z_sy_d)^2\big].$$

The first integral (A1) in this approximation gives the equation for determination of the  function
$\widetilde{t}(z)$:

$$cN{d\widetilde{t}(z)\over dz}-2\big[N_x^2+N_y^2+1\big]
\Big[{3({\bf M\ R})^2\over N^8\rho^8}+{{\bf M}^2\over N^6\rho^6}\Big]- \eqno(A9)$$
$$-\Big[N_x{d\widetilde{x}(z)\over dz} +N_y{d\widetilde{y}(z)\over dz}\Big]
+{2\over N^6\rho^6}\Big[{3({\bf M\ R})\over N^2\rho^2} \left({\bf R\ N}\right)-\left({\bf M\
N}\right)\Big]^2=0. $$ Before to integrate the equations (A7) and (A9), we make some estimations. According to
the statement of the problem, the source of hard emission
 is on the neutron star surface or in it
magnetosphere. Taking the neutron star radius equal to the 10 km, we can  assume that
$\sqrt{x_s^2+y_s^2+z_s^2}<10^2$ km. Detector of hard emission is on the near-Earth satellite. Usually, two
types of orbits are appropriate for such observations. First one is the low-altitude circular orbit, which
lays mainly under the Earth radiation belts. Its radius no more than approximately $7\cdot 10^3$ km. The other
type of an orbit is high-apogee elliptical one, on which satellite mainly is out of the Earth magnetosphere.
The apogee altitude for such orbits (INEGRAL space observatory for ex.)
 can be about $10^5$  ª¬.
Therefore, we can use approximations $|x_d|\leq10^5$  km and
 $|y_d|\leq10^5$ km for coordinates $x_d$ ¨ $y_d$.

The coordinate  $z_d$ value in the chosen coordinate system is the same as the distance from the pulsar to the
Earth. Because the nearest pulsar is on the distance of about a few kps, i.e.
 $\sim10^{17}$ km from the Earth, we  assume
that $z_d\sim 10^{17}$ km. Substituting these values in the expressions (A6), we obtain the
 following estimations:
$N_x\sim N_y\sim10^{-12},\  N_z=1.$ Therefore, we can simplify significantly the equations (A7) and (A9),
 keep there only asymptotically main terms
in the expansion on the small parameters $N_x$ ¨ $N_y\sim10^{-12}$.

In this case it is following from equations (A5) and (A8), that ${\bf R}=\{x_s,y_s,z\},$ $X=x_s,\ Y=y_s,\
N_x=N_y=0,\ N=1,\ p=0,\ q^2=x_s^2+y_s^2,\ \rho=\sqrt{z^2+q^2}.$ Thus, equations (A7) have the form:
$${d^2\widetilde{x}(z)\over dz^2}=
{180\over \rho^{12}}({\bf M\ R})^2x_s(x_s^2+y_s^2)+ {12\over \rho^{10}}({\bf M\ R})\Big[(3x_s^2
-5y_s^2) M_x-24({\bf M\ R})x_s+\eqno(A10)$$
$$+x_s(8y_sM_y+14zM_z)\Big]+{12\over \rho^8}\Big[6({\bf M\ R})M_x-x_s(2M_x^2+M_y^2+M_z^2)
-(y_sM_y+3zM_z)M_x\Big],$$
$${d^2\widetilde{y}(z)\over dz^2}=
{180\over \rho^{12}}({\bf M\ R})^2y_s(x_s^2+y_s^2)+{12\over \rho^{10}}({\bf M\ R})\Big[(3y_s^2
-5x_s^2) M_y-24({\bf M\ R})y_s+$$
$$+y_s(8x_sM_x+14zM_z)\Big]+{12\over \rho^8}\Big[6({\bf M\ R})M_y-y_s(2M_y^2+M_x^2+M_z^2)
-(x_sM_x+3zM_z)M_y\Big].$$

The first integral (A9) also becomes more simple in this case
$$c{d\widetilde{t}(z)\over dz}-2\Big[{3({\bf M\ R})^2\over \rho^8}+
{{\bf M}^2\over \rho^6}\Big] +{2\over \rho^6}\Big[{3({\bf M\ R})\over \rho^2}z - M_z\Big]^2=0.\eqno(A11) $$

To solve equations (A10), we obtain:
$$\widetilde{x}(z)= {3\over  64q^9}\hbox{atan}\Big( {z\over  q}\Big)
\big[2M_xM_zq^4-175zx_s(M_xx_s+M_yy_s)^2+\eqno(A12)$$
$$+10q^2(5M_zx_s-3zM_x)(M_xx_s+M_yy_s)
-5zx_sq^2(16M_y^2+15M_z^2)+$$
$$+80y_szq^2M_xM_y\big]+ {1\over  64\rho^2q^6}\big[
30(5x_szM_z+M_xq^2)(M_xx_s+M_yy_s)+$$
$$+2q^2M_x(3M_zz-40M_yy_s)
+5x_sq^2(16M_y^2+15M_z^2)+175x_s(M_xx_s+$$
$$+M_yy_s)^2\big]+ {1\over  32\rho^4q^4}
\big[x_sq^2(16M_y^2+15M_z^2)+(6M_xq^2+50x_szM_z)\times$$
$$\times(M_xx_s+M_yy_s)+2q^2M_x(M_zz-8M_yy_s)+35x_s(M_xx_s+M_yy_s)^2\big]+$$
$$+ {5\over 8\rho^6q^2}\big[
x_s(M_xx_s+M_yy_s)^2+q^2(2M_zM_xz-M_z^2x_s)+2(M_xq^2+$$
$$+M_zx_sz)(M_xx_s+M_yy_s)
\big]-{9x_s({\bf M\ R})^2\over 4\rho^8} ,$$
$$\widetilde{y}(z)= {3\over  64q^9}\hbox{atan}\Big( {z\over  q}\Big)
\big[2M_yM_zq^4-175zy_s(M_xx_s+M_yy_s)^2+$$
$$+10q^2(5M_zy_s-3M_yz)(M_xx_s+M_yy_s)
-5zy_sq^2(16M_x^2+15M_z^2)+$$
$$+80x_szq^2M_xM_y\big]+ {1\over  64\rho^2q^6}\big[
30(5M_zy_sz+M_yq^2)(M_xx_s+M_yy_s)-$$
$$+2q^2M_y(3M_zz-40M_xx_s)
+5y_sq^2(16M_x^2+15M_z^2)+175y_s(M_xx_s+$$
$$+M_yy_s)^2\big]+ {1\over  32\rho^4q^4}
\big[y_sq^2(16M_x^2+15M_z^2)+(6M_yq^2+50M_zy_sz)\times$$
$$\times(M_xx_s+M_yy_s)+2q^2M_y(M_zz-8M_xx_s)+35y_s(M_xx_s+M_yy_s)^2\big]+$$
$$+ {5\over 8\rho^6q^2}\big[
y_s(M_xx_s+M_yy_s)^2+q^2(2M_zM_yz-M_z^2y_s)+$$
$$+2(M_yq^2+M_zy_sz)(M_xx_s+M_yy_s)
\big]-{9y_s({\bf M\ R})^2\over 4\rho^8}.$$

Integrating the equation (A11), we find the explicit dependence $\widetilde{t}(z)$, i.e. the law of a hard
emission pulse move along beams  from it
 common source to the detector:
$$\widetilde{t}(z)=\big[25(M_xx_s+M_yy_s)^2+q^2(16{\bf M}^2-M_z^2)\big]
\Big[ {z(3\rho^2+2q^2)\over  64\rho^4q^6}+{3\over  64q^7}\hbox{atan}\Big( {z\over  q}\Big)\Big]+\eqno(A13)$$
$$+{1\over 8\rho^6q^2}\big[5z(M_xx_s+M_yy_s)^2+16q^2M_z(M_xx_s+M_yy_s)+3M_z^2q^2z\big]+
{9\over 4\rho^8}\big[z(M_xx_s+M_yy_s)^2-$$
$$-q^2M_z(2M_xx_s+2M_yy_s+M_zz)\big].$$

It is necessary to note, that expressions (A12) and (A13) are the
 partial solutions of non-uniform equations (A10)
and (A11). From the mathematics rules, they should be added by the
 general solutions of corresponding uniform equations, i.e. make
 replacement
$$\widetilde{t}(z)\to \widetilde{t}(z)+a_0,\ \
\widetilde{x}(z)\to \widetilde{x}(z)+a_1+b_1z,\ \ \widetilde{y}(z)\to \widetilde{y}(z)+a_2+b_2z,$$ where
$a_0,\ a_1,\ a_2,\ b_1,\  b_2$ are the arbitrary constants.

However, if to substitute these general solutions (A10) and (A11) in
 the expressions (A3) then all constants
$a_0,\ a_1,\ a_2,\ b_1,\  b_2$ exactly cancel.  Therefore, further we will use more simple expressions
 (A12) and (A13).



\end{document}